\documentstyle[preprint,epsf,aps,floats,tighten]{revtex} % for preprint mode with epsf
\def\met{\mbox{${\hbox{$E$\kern-0.6em\lower-.1ex\hbox{/}}}_T~$}} %missing ET
\def\vmet{\mbox{${\hbox{$\vec E$\kern-0.6em\lower-.1ex\hbox{/}}}_T~$}} %missing ET
\def\D0{D\O}                            %D0
\begin{document}

%\preprint{FERMILAB-Pub-97/109-E}
%
% ======> Title of the paper goes here <====================
%
\title{D\O\ Results on $W$ Boson Properties}

% ====> Use the following for the Lepton-Photon symposium in Hamburg <=====
%\author{\centerline{The D\O\ Collaboration
%  \thanks{Submitted to the {\it XVIII International Symposium on Lepton Photon
%          Interactions,}
%          \hfill\break
%          July 28 -- August 1, 1997, Hamburg, Germany.}}}
%
% ====> Use the following for the HEP97 conference in Jerusalem <=====
%\author{\centerline{The D\O\ Collaboration
%  \thanks{Submitted to the {\it International Europhysics Conference 
%          on High Energy Physics,}
%          \hfill\break
%          August 19 -- 26, 1997, Jerusalem, Israel.}}}
%
% ====> Use the following for the HCP 97 Proceedings <=====
\author
 {
  \centerline{
  Kathleen Streets\thanks{Presented at the 
  {\it Hadron Collider Physics XII} Conference, \hfill\break 
  \hspace*{0.3cm} June 5 -- 11, 1997, Stony Brook, New York, USA.}
}
}

% for HCP june97
\address{
  \begin{center}
   {\rm (for the D\O\ Collaboration)}\\
  New York University, NY, NY 10003, USA
  \end{center}
}

%\address{
%\centerline{Fermi National Accelerator Laboratory, Batavia, Illinois 60510}
%}   

%
% Indicate today's date
%
\date{\today}

\maketitle

%
% ==============> Text of the abstract goes here <=====================
% 
\begin{abstract}

The D\O\ experiment collected 
  $\approx 15 ~\rm{pb^{-1}}$ in run 1A  (1992-1993) 
  and $\approx 89 ~\rm{pb^{-1}}$ in run 1B (1994-1995)
  of the Fermilab Tevatron Collider using $p\bar p$ 
  collisions at $\sqrt s = 1.8 ~\rm{TeV}$.
Results from analyses of events with $W$ and $Z$ bosons are
  presented for the run 1B data samples.
From  $W\rightarrow e\nu,\mu\nu$
  and $Z\rightarrow ee,\mu\mu$ decays, the $W$ and $Z$ production 
  cross sections and the $W$ width  are determined.
Events with  $W\rightarrow \tau\nu$ decays are used to
  determine the ratio of the electroweak gauge coupling constants
  as a measure of lepton  universality.
Using $W\rightarrow e\nu$ and $Z\rightarrow ee$ decays,
  the  $W$ boson mass is measured.
                             
\end{abstract}

%\pacs{PACS numbers 14.65.Ha, 13.85.Qk, 13.85.Ni}

%=======================================================================
%\vskip 1cm
%\newpage
%\begin{center}
%\input{LIST_OF_AUTHORS.TEX}
%\end{center}

\normalsize

\vfill\eject

%==========================================================================

\section{Introduction}

The D\O\ experiment collected $\approx 15 ~\rm{pb^{-1}}$ in run 1A
  (1992-1993)   and $\approx 89 ~\rm{pb^{-1}}$ in run 1B (1994-1995)
  of the Fermilab  Tevatron 
   Collider using $p\bar p$ collisions at $\sqrt s = 1.8 ~\rm{TeV}$.
Results are presented from data collected by the D\O\ experiment
  that test the Standard Model (SM) of electroweak interactions\cite{ewsm}.
Measurements of the $W$ and $Z$ boson
  production cross sections,  the $W$ decay width, the ratio of the
  gauge coupling constants,  and the $W$ mass are presented.

%==========================================================================

\section{The D\O\ Detector}

The D\O\ detector was designed to study a variety of high
  transverse momentum ($p_T$) physics
  topics and has been described in   detail elsewhere \cite{D0detector}.
It does not have a central magnetic field, making possible a
  compact, hermetic detector with almost full solid angle coverage.
The detector has an inner tracking system  which measures
  charged tracks to a pseudo-rapidity $\eta  <3.2$, where $\eta = -\ln \tan
  {\theta \over 2}$ and  $\theta$ is the polar angle.
The tracking system is surrounded by  finely-segmented uranium liquid-argon
  calorimeters  (one central and  two end-caps).
Surrounding the calorimeter is a muon magnetic spectrometer which consists of 
  magnetized iron toroids that are situated between the first two 
  of three layers of  proportional drift tubes.

Electrons and photons were identified by the shape of their
  energy deposition in the calorimeter and a matching track (for electrons).
The energy ($E$) was measured by the calorimeter with a
  resolution of $\approx 15\% / \sqrt{E} ~{\rm(GeV)}$.
%The muon momentum resolution is $\sigma(1/p) = 0.18(p-2)/p^2 \oplus 0.008$
%  ($p$ in GeV/c).
Neutrinos were not identified in the detector but their
  transverse momentum  was inferred from the missing transverse energy in the
  event:
  \vmet $ = -\sum_i E_{i} \sin \theta$,
  where the sum $i$ extends over all cells in the calorimeter.
Muons were identified by a track in the muon chambers matched with a
  track in the central tracking chambers.

%==========================================================================

\section{$W$ and $Z$ Production}

Events in which a $W$ or $Z$ boson is produced are used to measure
 the cross section times branching fraction, the $W$ width and
 the ratio of the gauge couplings.
In these analyses, the $W$ and $Z$ gauge bosons
  are identified through their leptonic decay modes:
  $W\rightarrow e\nu,\mu\nu,\tau\nu$ and $Z\rightarrow ee,\mu\mu$.
These modes have a cleaner signature and are  easier to distinguish
  from the background of QCD multijet production than hadronic decay modes.
The events with decays into $e$'s and $\mu$'s are selected 
  by requiring a high-$p_T$ $e$ or $\mu$  and large \met  for  $W$'s and two 
  high-$p_T$ $e$'s or $\mu$'s for $Z$'s.
The hadronic decay of the $\tau$ is used to to select the 
  $W\rightarrow \tau \nu$ events.

\subsection{Production Cross Sections}

The measurement of the product of the cross section and the branching 
  fraction for $W$'s and $Z$'s provides a fundamental test of 
  the Standard Model.
These measurements have been published for the run 1A data 
  sample\cite{wzxsect1a} and the preliminary results
  are presented here for the run 1B data sample.

\begin{figure}[h]
 \begin{center}
  \begin{tabular}{cc}
    \epsfxsize = 7cm   \epsffile{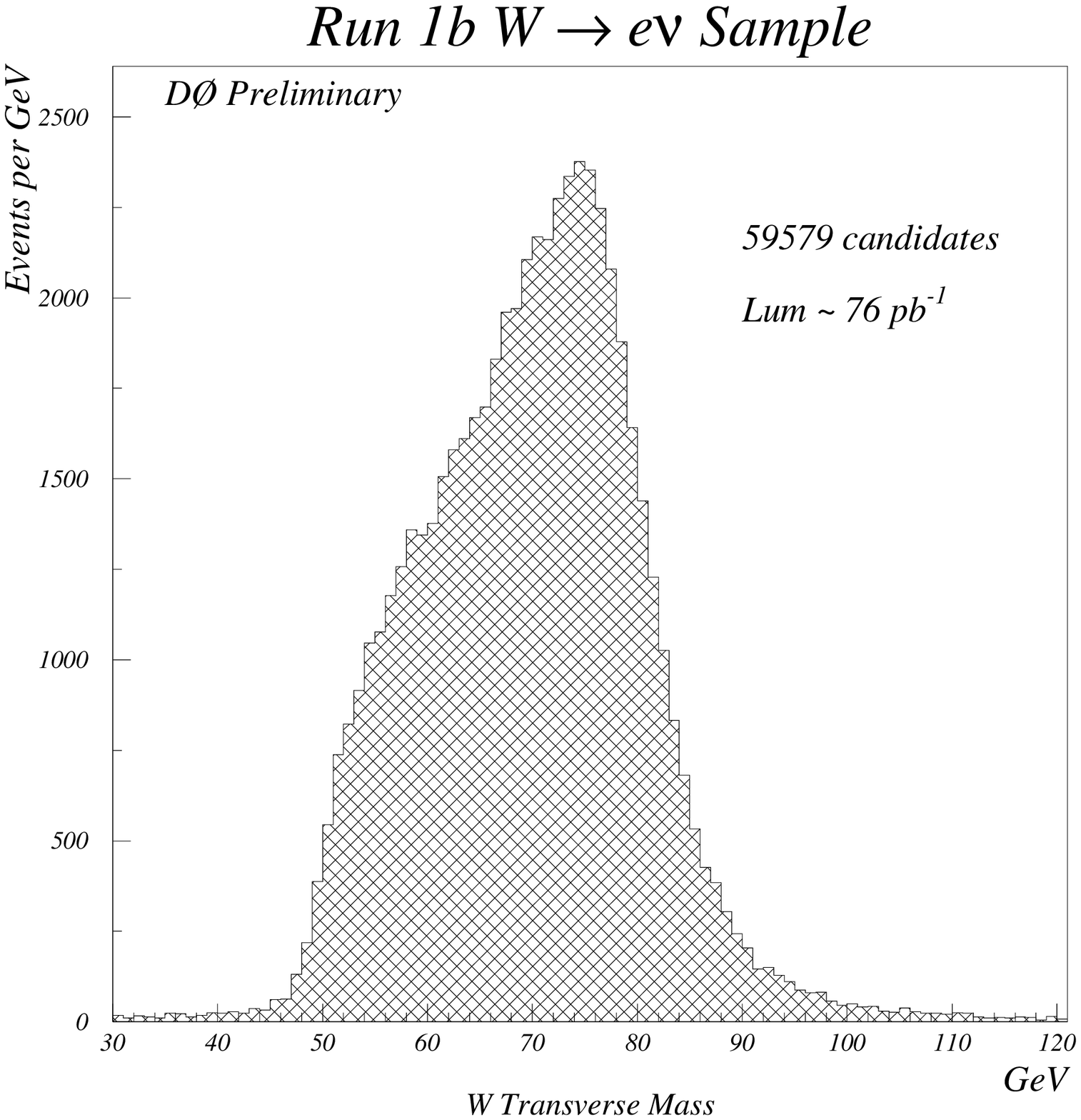} &
    \epsfxsize = 7cm   \epsffile{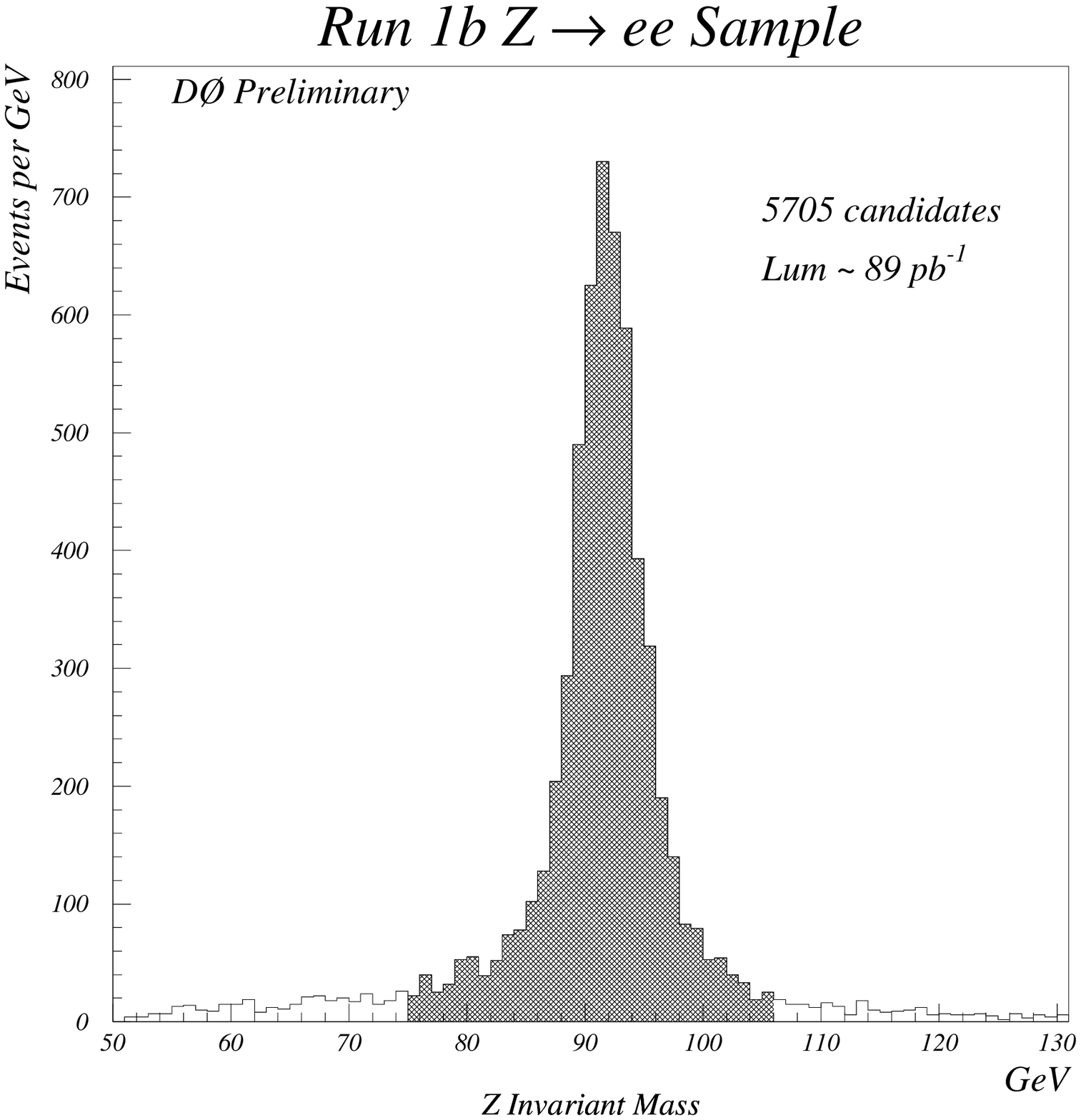} \\
    \epsfxsize = 7cm   \epsffile{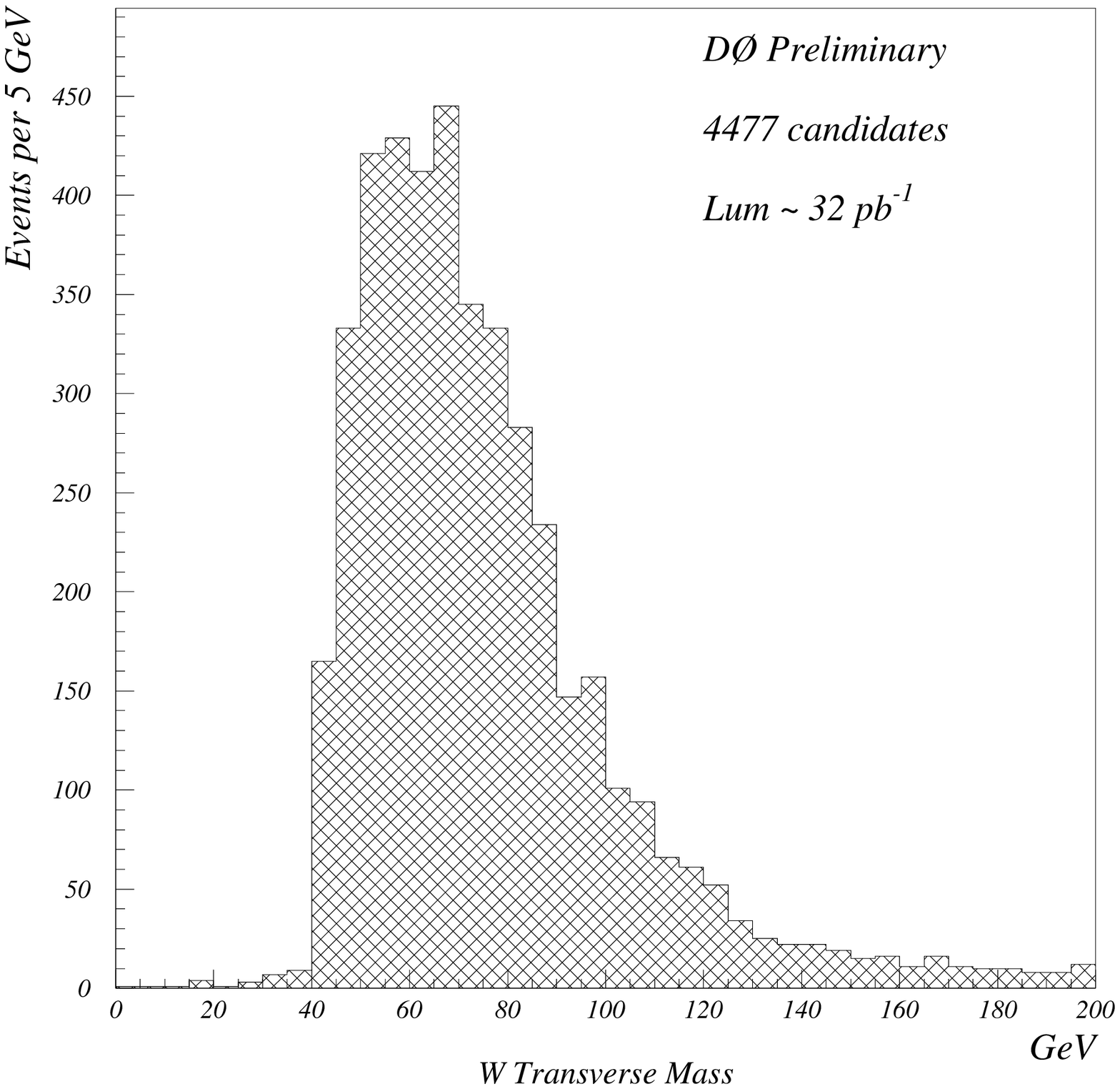} &
    \epsfxsize = 7cm   \epsffile{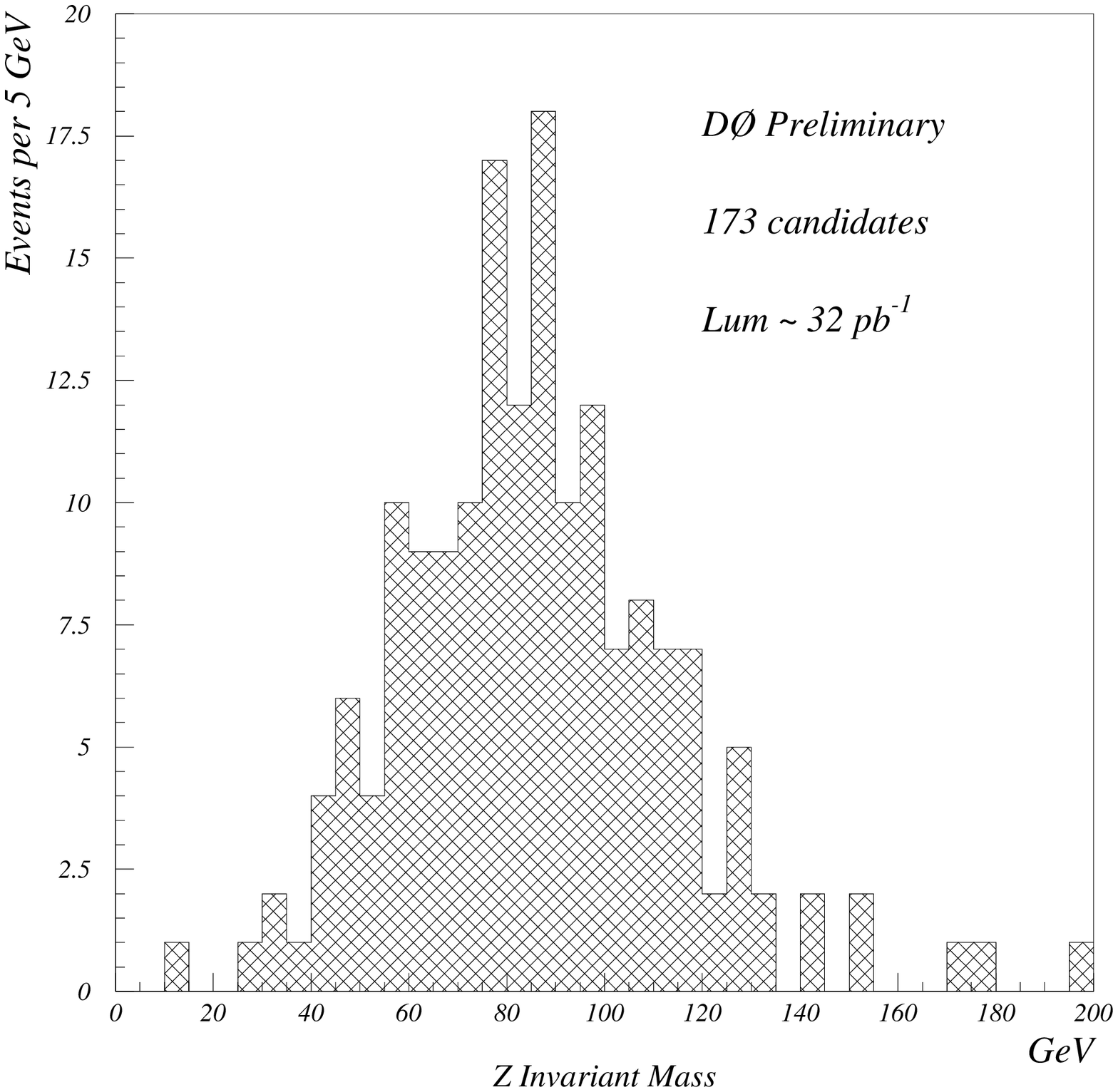} \\
  \end{tabular}
\caption{Transverse and invariant mass distributions for the 
         $W\rightarrow e\nu,\mu\nu$ and $Z\rightarrow ee,\mu\mu$ 
          run 1B data samples.}
\label{FIG:FIG1}
\end{center}
\end{figure}

For the final event selection in this analysis, 
  electrons  were restricted to a region $|\eta|<1.1$ and
  $1.5<|\eta|<2.5$ and muons to a region $|\eta| <1.0$.
The $W\rightarrow e\nu$ events were selected by requiring the transverse
  energy of the electron  $E_T > 25 ~\rm {GeV}$
  and \met $ > 25 ~\rm {GeV}$ and the  $Z\rightarrow ee$ events
  were required to have two $e$'s   with $E_T > 25 ~\rm {GeV}$.
The $W\rightarrow \mu \nu$ event selection required
   $p_T(\mu) > 20 ~\rm {GeV},$ and \met $ > 20 ~\rm {GeV}$ and the
   $Z\rightarrow \mu\mu$ selection required
   $p_T > 15,20 ~\rm {GeV}$ for the two $\mu$'s.
The transverse mass for $W$ events and invariant mass for $Z$ events
  in the final data samples are shown in Fig. \ref{FIG:FIG1}.
Table \ref{TABLE:TAB1} gives the number of events observed, the
  acceptance, the efficiency, the background and the luminosity for these
  data samples.

\begin{table}[h]
\begin{center}
\begin{tabular}{lcc}
 & Electron & Muon \\
\hline 
\# $W$ candidates   & 59579          & 4472 \\
Acceptance $(\%)$ & $43.4 \pm 1.5$ & $20.1 \pm 0.7$ \\
$\epsilon_W(\%)$ & $70.0 \pm 1.2$ & $24.7 \pm 1.5$ \\
Background$(\%)$ & $8.1 \pm 0.9$  & $18.6 \pm 2.0$ \\
Luminosity $(pb^{-1})$ & $75.9 \pm 6.4$ & $32.0 \pm 2.7$ \\
\hline 
\# $Z$ candidates   & 5702           & 173 \\
$A_Z(\%)$        & $34.2 \pm 0.5$ & $5.7 \pm 0.5$  \\
$\epsilon_Z(\%)$ & $75.9 \pm 1.2$ & $43.2 \pm 3.0$ \\
Bkg.$(\%)      $ & $4.8 \pm 0.5$  & $8.0 \pm 2.1 $ \\
Lum. $(pb^{-1})$ & $89.1 \pm 7.5$ & $32.0 \pm 2.7$ \\
\end{tabular}
\caption{The quantities used to measure the preliminary cross sections for 
         $W\rightarrow e\nu,\mu\nu$ and $Z\rightarrow ee,\mu\mu$ for the
         run 1B data sample.}
\label{TABLE:TAB1}
\end{center}
\end{table}
                            
\begin{figure}[h]
 \begin{center}
  \begin{tabular}{c}
    \epsfxsize = 10cm   \epsffile{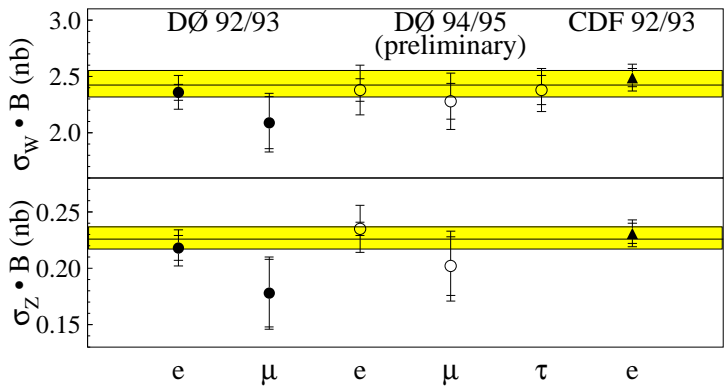}
  \end{tabular}
\caption{Tevatron measurements for the cross sections times branching ratios
         for $W\rightarrow e\nu,\mu\nu,\tau\nu$ and $Z\rightarrow ee,\mu\mu$
         compared to SM predictions.}
\label{FIG:FIG2}
\end{center}
\end{figure}

The preliminary measurements of the cross section times branching fraction 
  ($\sigma \cdot B$) are given in table \ref{TABLE:TAB2}  and are shown in
  Fig. \ref{FIG:FIG2} along with the results from CDF\cite{wzxsectcdf}.
The $\tau$ results shown will be discussed in section \ref{sec:tau}.
Also shown in Fig. \ref{FIG:FIG2} are comparisons of $\sigma \cdot B$
  with SM predictions\cite{wzsmpred}.
The predictions use the CTEQ2M parton distribution functions 
  (pdf)\cite{cteq2m}.

\begin{table}[h]
\begin{center}
 \begin{tabular}{llllll}
   $\sigma_W \cdot B(W\rightarrow e\nu)$   & = &
      2.38 & $\pm$ 0.01 & $\pm$ 0.09 & $\pm$ 0.20 nb \\
   $\sigma_W \cdot B(W\rightarrow \mu\nu)$ & = &
      2.32 & $\pm$ 0.04 & $\pm$ 0.16 & $\pm$ 0.19 nb \\
   $\sigma_Z \cdot B(Z\rightarrow ee)$     & = &
      0.235 & $\pm$ 0.003 & $\pm$ 0.005 & $\pm$ 0.020 nb \\
   $\sigma_Z \cdot B(Z\rightarrow \mu\mu)$ & = &
      0.202 & $\pm$ 0.016 & $\pm$ 0.020 & $\pm$ 0.017 nb \\
    ~ & ~ & ~ & $(\pm {\rm stat})$ & $(\pm {\rm syst})$ &
                    $(\pm {\rm lum})$ \\
  \end{tabular}
\caption{The preliminary cross sections for 
         $W\rightarrow e\nu,\mu\nu$ and $Z\rightarrow ee,\mu\mu$.}
\label{TABLE:TAB2}
\end{center}
\end{table}

\subsection{$W$ width}

The ratio of the $W$ and $Z$ production cross sections can be used to
  measure the leptonic branching ratio $B(W\rightarrow l\nu)$ and extract the
  $W$ width ($\Gamma_W$).
From the measured width, a limit may be placed on unexpected decay 
  modes of the $W$.
Many common systematic errors, including the luminosity error,
  cancel in the leptonic branching ratio:
  $$ R = {{\sigma_W \cdot B(W\rightarrow l\nu)}\over
          {\sigma_Z \cdot B(Z\rightarrow ll)}}  =
              {{\sigma_W} \over {\sigma_Z}}
              {{\Gamma(W\rightarrow l\nu)}\over{\Gamma (Z\rightarrow ll)}}
              {{\Gamma_Z}\over{\Gamma_W}}.$$
Using the results above for  $\sigma \cdot B$ and combining the electron
  and muon measurements, we obtain a preliminary run 1B
  result of  $R = 10.32 \pm 0.43 $.
The leptonic branching fraction of the $W$ may then be calculated,
  $ B(W\rightarrow l\nu) = B(Z\rightarrow ll) \cdot 
                   {{(\sigma_Z}/{\sigma_W})} \cdot R = (10.43\pm 0.44)\%$
  using the measured value of $R$,
  the value of   $B(Z\rightarrow ll)$  from LEP measurements\cite{lepbrzll}
  and  $\sigma_W/ \sigma_Z = 3.33 \pm 0.03$ from 
   the  SM prediction\cite{smpred}.
The total width of the W is then obtained from this measurement of
  $ B(W\rightarrow l\nu)$ and the value of $\Gamma(W\rightarrow l\nu)$
  from SM predictions\cite{smwwidth}.
The preliminary run 1B measurement is  
  $$\Gamma_W = 2.159\pm 0.092~ \rm{GeV}.$$

Comparison of the published world average 
  $\Gamma_W = 2.062\pm 0.059 ~\rm{GeV}$\cite{wzxsect1a}
   (does not include the run 1B measurement) 
  with the SM prediction  
    $\Gamma_W = 2.077\pm 0.014 ~\rm{GeV}$\cite{smwwidth},
  gives a $95\%$ confidence level
  upper limit of $\Delta \Gamma_W < 109 ~\rm {MeV}$  on
  unexpected (non-SM) decays of the $W$.
                                            
\subsection{Measurement of the Ratio of the Couplings}
\label{sec:tau}

The decay $W\rightarrow \tau \nu$ is studied as a test of lepton universality
  by measuring the ratio of the electroweak coupling 
   constants  $g_{\tau}^W/g_e^W$.
The $\tau$ events are obtained from a sample in which inelastic
  collisions were selected by requiring a single interaction signature from 
  the Level 0 trigger.
The integrated luminosity for the $\tau$ trigger  used in this analysis is
  $16.8\pm 0.9~{\rm pb^{-1}}$.

\begin{figure}[h]
 \begin{center}
  \begin{tabular}{c}
    \epsfxsize = 10cm   \epsffile{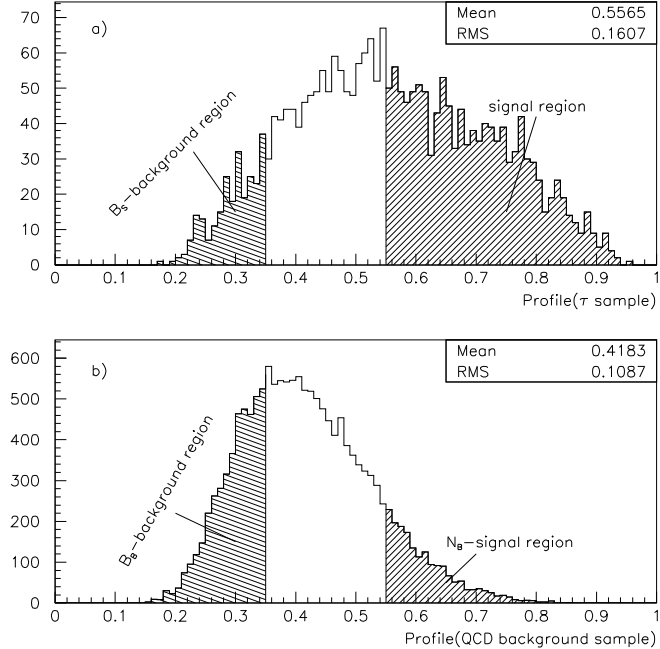}
  \end{tabular}
\caption{The distribution of the profile quantity for the $\tau$ 
   candidate sample    and for the QCD background sample.}
\label{FIG:tau1}
\end{center}
\end{figure}

To select the $W\rightarrow \tau\nu$ events from the $W$ data sample, 
  the hadronic decay of the $\tau$ is used.
These events are identified by the presence of an isolated, narrow jet.
Jets were reconstructed using a cone algorithm with radius $0.7$ 
  in $\eta - \phi$ space
  and the width of the jet was required to be $rms_{jet}<0.25$.
The requirements that $E_T$(jet)$>25$ GeV, ($|\eta|<0.9$),  $\met >25$ GeV
  and that there be no opposite jet  were placed on the data sample.
In order to separate the events with a jet from a $\tau$ decay from the
  large background of QCD jets, the profile distribution of the
  jets is used.
The profile is defined as the sum of the highest two tower $E_T$'s 
  divided  by the cluster $E_T$.
The profile distributions from the $\tau$ sample and the QCD background
  sample (selected from events with low $\met$) are shown 
  in Fig. \ref{FIG:tau1}.
A requirement that the profile variable be $> 0.55 $ is made to
  select the final $\tau$ event sample.
The shaded low-profile region in Fig. \ref{FIG:tau1}a is used to
  estimate the remaining QCD background.

The number of signal events contained in the final data sample is listed
  in table \ref{TABLEtau1} along with the estimated background contributions.
The preliminary value of the cross section times branching ratio is
  $\sigma\cdot B(W\to\tau\nu)=2.38\pm0.09(stat)\pm0.10(syst)~{\rm nb}$
  where the error due to the luminosity has not been included.
Comparing this value with the measurement of 
  $\sigma\cdot B(W\to e\nu)$ from run 1A\cite{wzxsect1a},
%=2.38 \pm 0.01(stat) \pm 0.09 (syst)~{\rm nb}$,
  the ratio of the couplings is determined
  $g_{\tau}^W/g_e^W=1.004\pm0.019(stat)\pm0.026(syst)$.
This measurement shows good agreement with $e-\tau$ universality 
  at high energy.

\begin{table}[h]
\begin{center}
\begin{tabular}{rc}
  & Number of Events \\ 
 Final Data Sample & 1202 \\
QCD Background & $106\pm7\pm5$ \\
Noise Events & $81\pm14$ \\
$Z\to\tau\tau$  & $32\pm5$ \\
$W\to e\nu$ & $3\pm1$ \\ 
\end{tabular}
\caption{The quantities used to measure the run 1B
  preliminary cross section for $W\rightarrow \tau\nu$.}
 \label{TABLEtau1}
\end{center}
\end{table}

%==========================================================================

\section{$W$ Mass}

\begin{figure}[h]
 \begin{center}
\vskip -0.5in
  \begin{tabular}{cc}
    \epsfxsize =  8.5cm   \epsffile{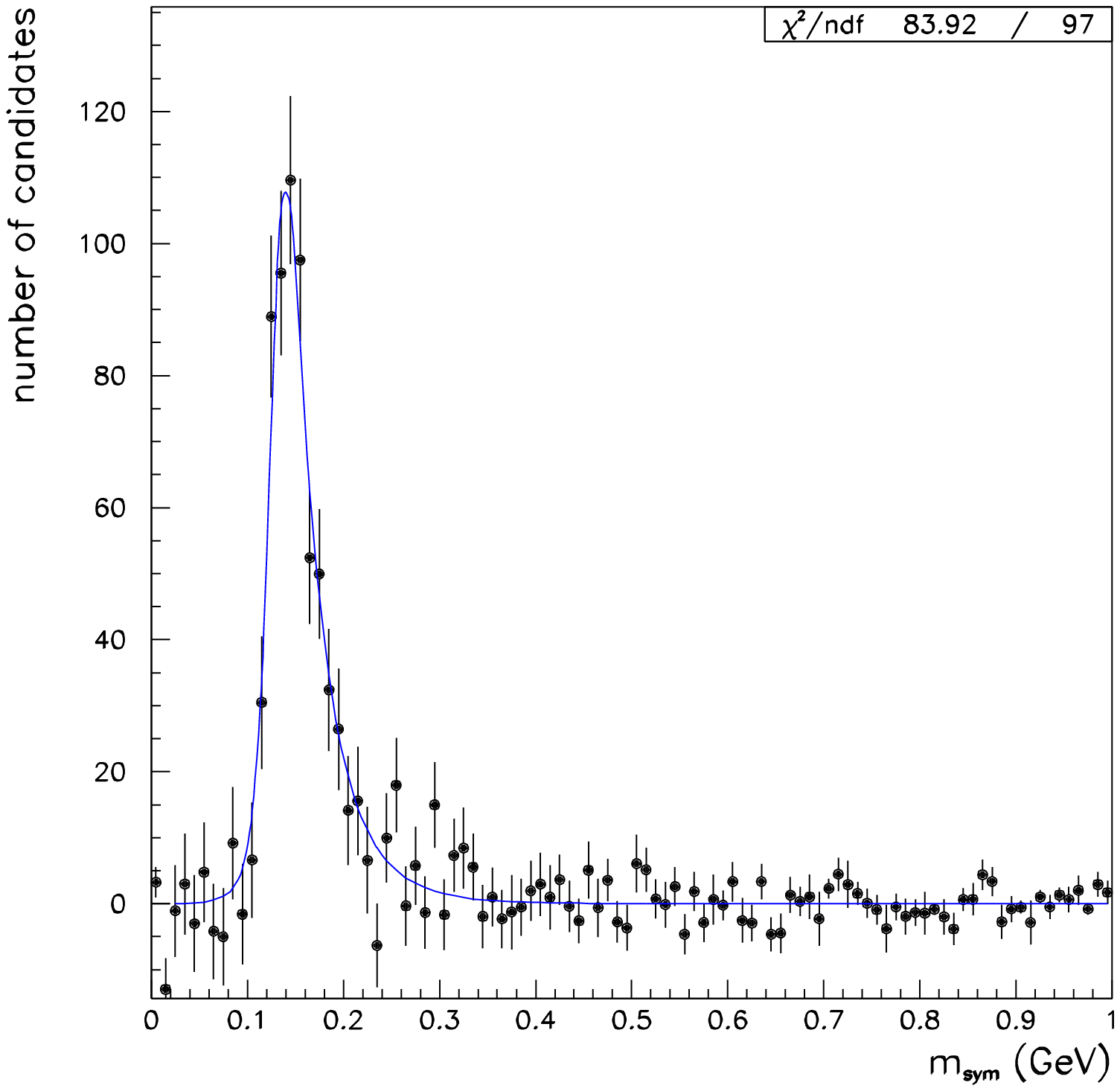} &
    \epsfxsize =  8.5cm   \epsffile{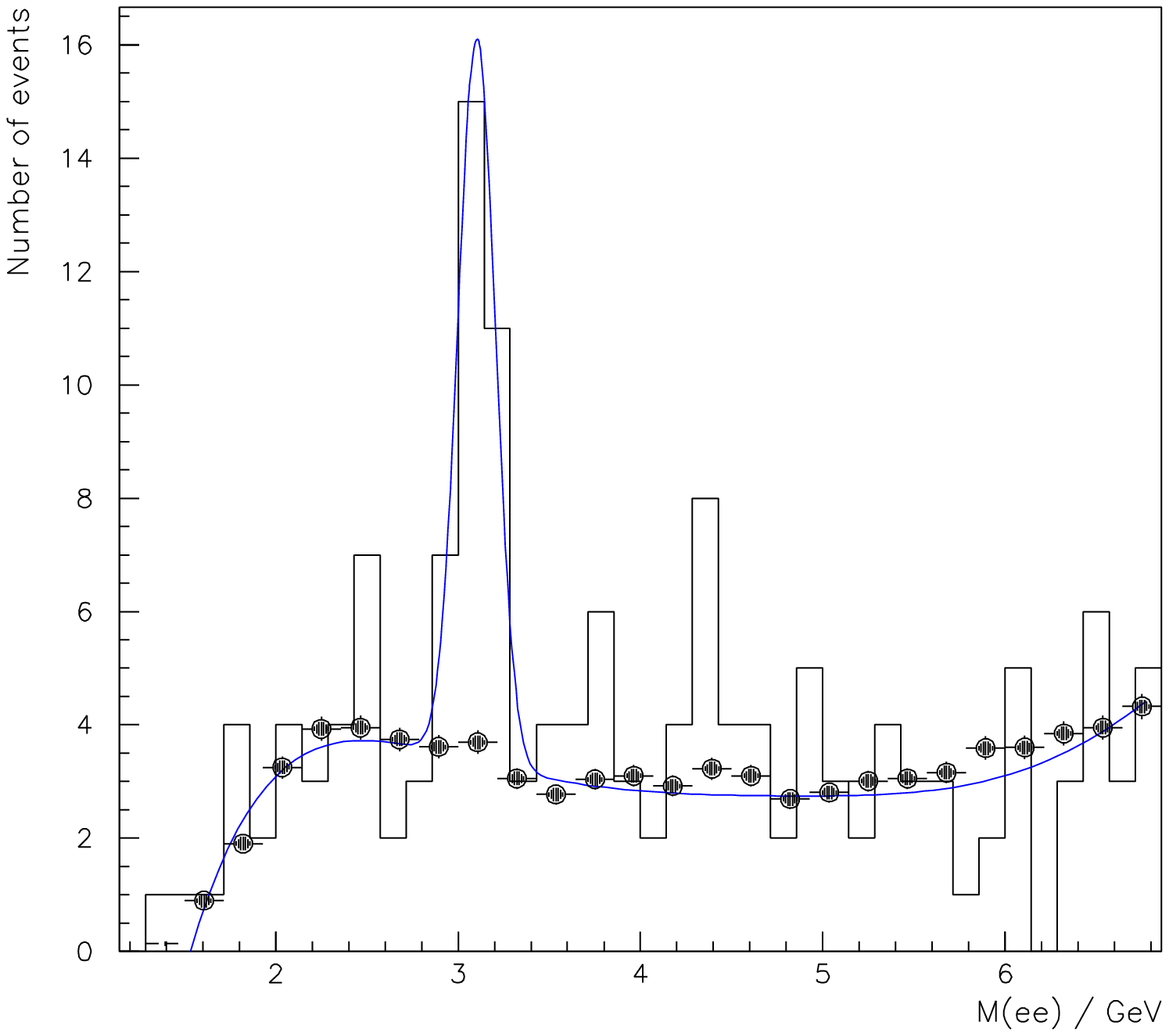} \\
  \end{tabular}
\vskip -1.2in
  \begin{tabular}{cc}
    \epsfxsize =  7.0cm   \epsffile{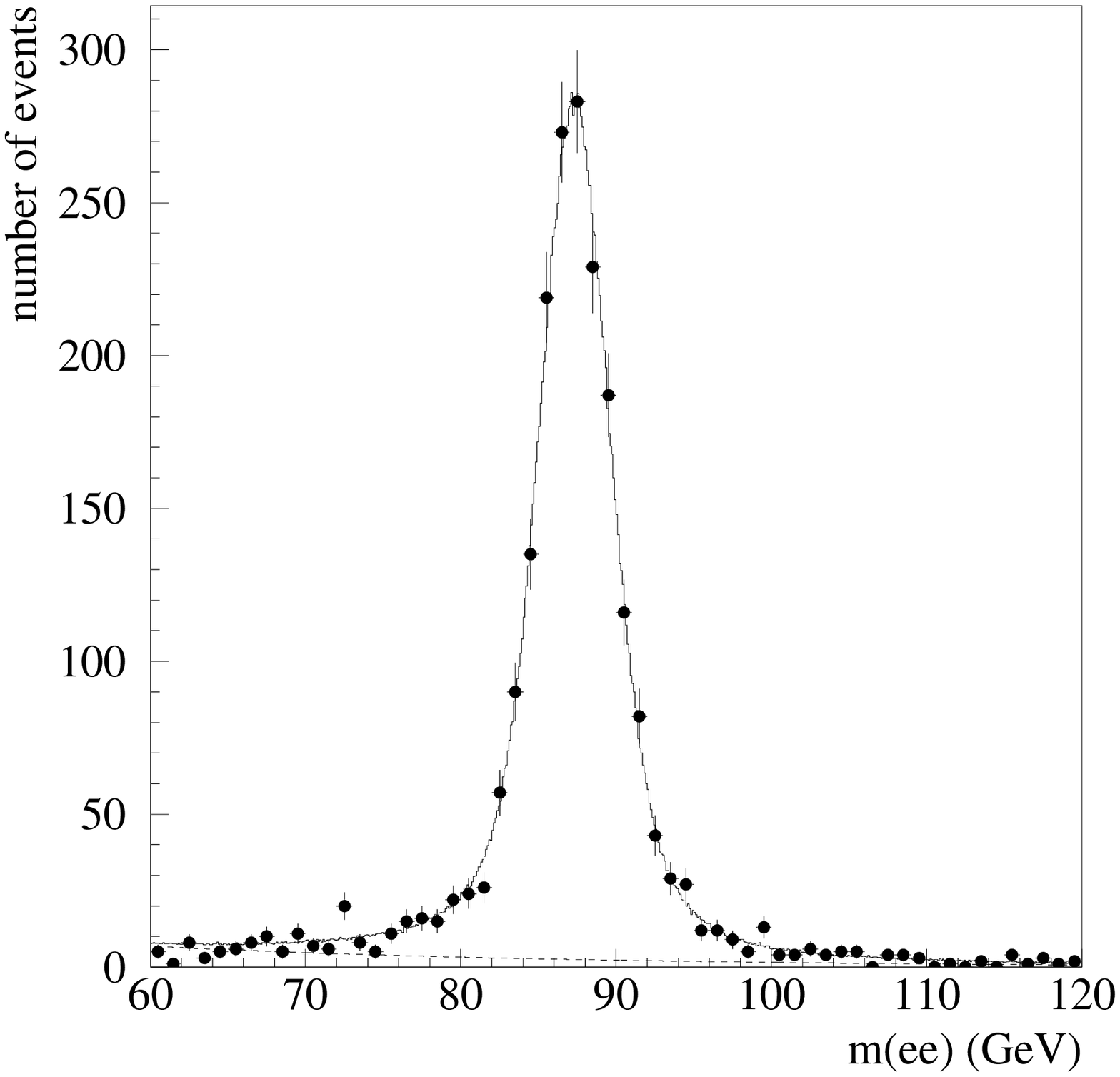} & 
    \epsfxsize =  7.0cm   \epsffile{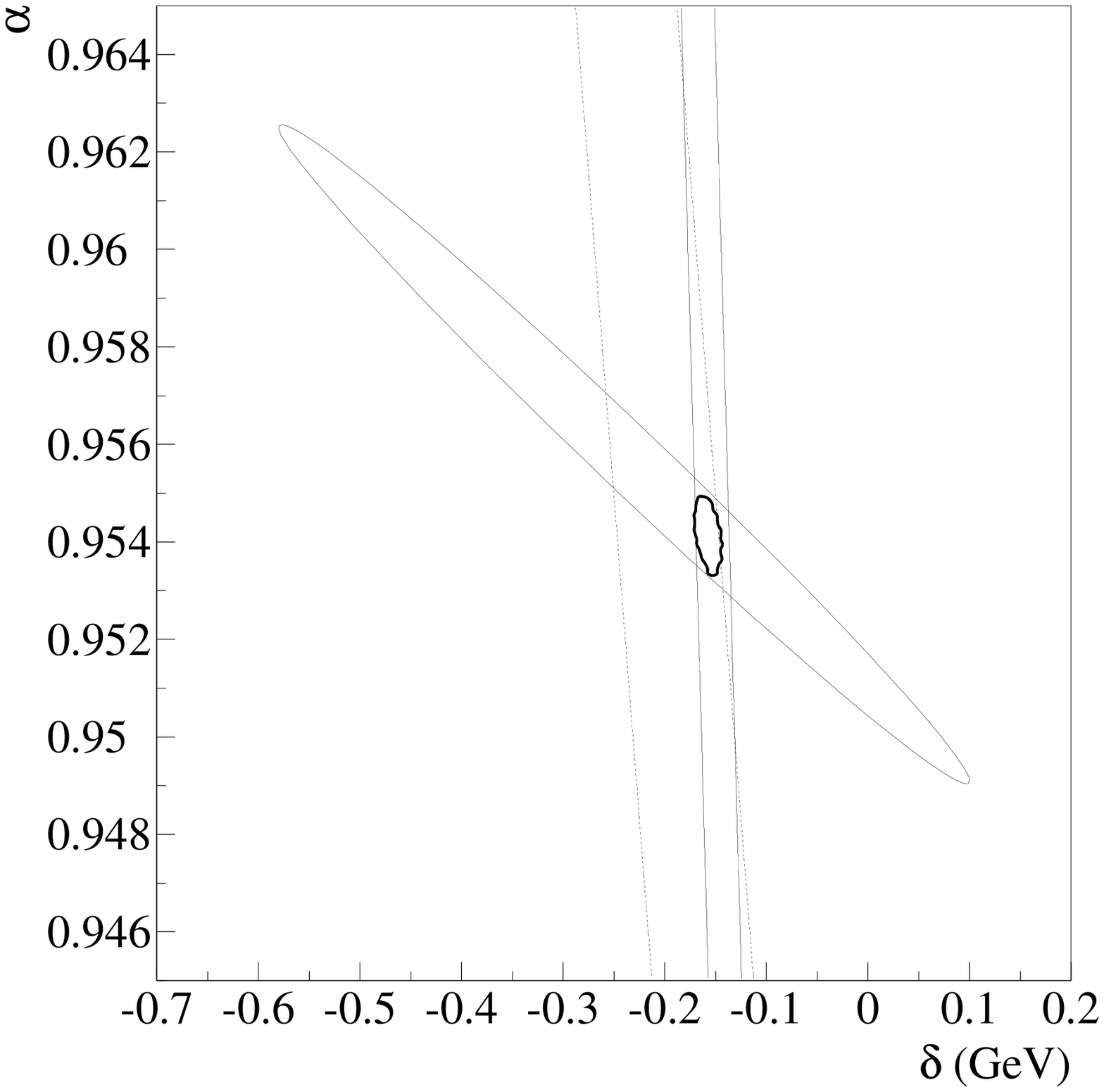}\\
  \end{tabular}
%\vspace{1.0in}
 \end{center}
  \caption{
     The invariant mass distributions are shown for 
     (a) $\pi^0 \rightarrow \gamma\gamma \rightarrow e^+e^-e^+e^-$ 
          events (points),
     (b) $J/\psi \rightarrow ee$ events (points), and
     (c) $Z \rightarrow ee$ events (points), compared to 
         Monte Carlo simulations (line).
     (d) Constraints on $\alpha$ and $\delta$ from
        $Z \rightarrow ee$  decays (large ellipse),
        $J/\psi \rightarrow ee$ decays (wide band), 
        $\pi^0 \rightarrow \gamma\gamma \rightarrow e^+e^-e^+e^-$ 
        decays (narrow band),
        and for all three combined (small ellipse).}
\label{FIG:w1}
\end{figure}

The electroweak Standard Model can be specified by three parameters.
These may be taken to be $\alpha$, $G_F$
  and $M_Z$,  all measured to $< 0.01\%$.
At lowest order, the $W$ mass is precisely defined as
  $M_W = A/\sin\theta_W$ with $\sin^2\theta_W = 1- M_W^2/M_Z^2$, where
  $\theta_W$ is the weak mixing angle and $ A = (\pi\alpha / \sqrt 2 G_F)^{1/2}$.
The current data are sufficiently precise to require comparison to theoretical
  predictions which include higher order corrections.
These corrections have contributions due to the
  running of $\alpha \rightarrow \alpha(M_Z^2)$ and to loop diagrams which
  introduce a dependence on the square of the top quark mass, 
   $m_{top}$, and the log of  the Higgs mass, $M_{H}$.
A precision measurement of the $W$ mass therefore defines the
  size of the radiative corrections in the SM and along with $m_{top}$ it
  can constrain $M_{H}$.
A direct measurement also serves to test the consistency of the SM.
               
\begin{figure}
 \begin{center}
  \begin{tabular}{cc}
    \epsfxsize = 7.0cm \epsffile{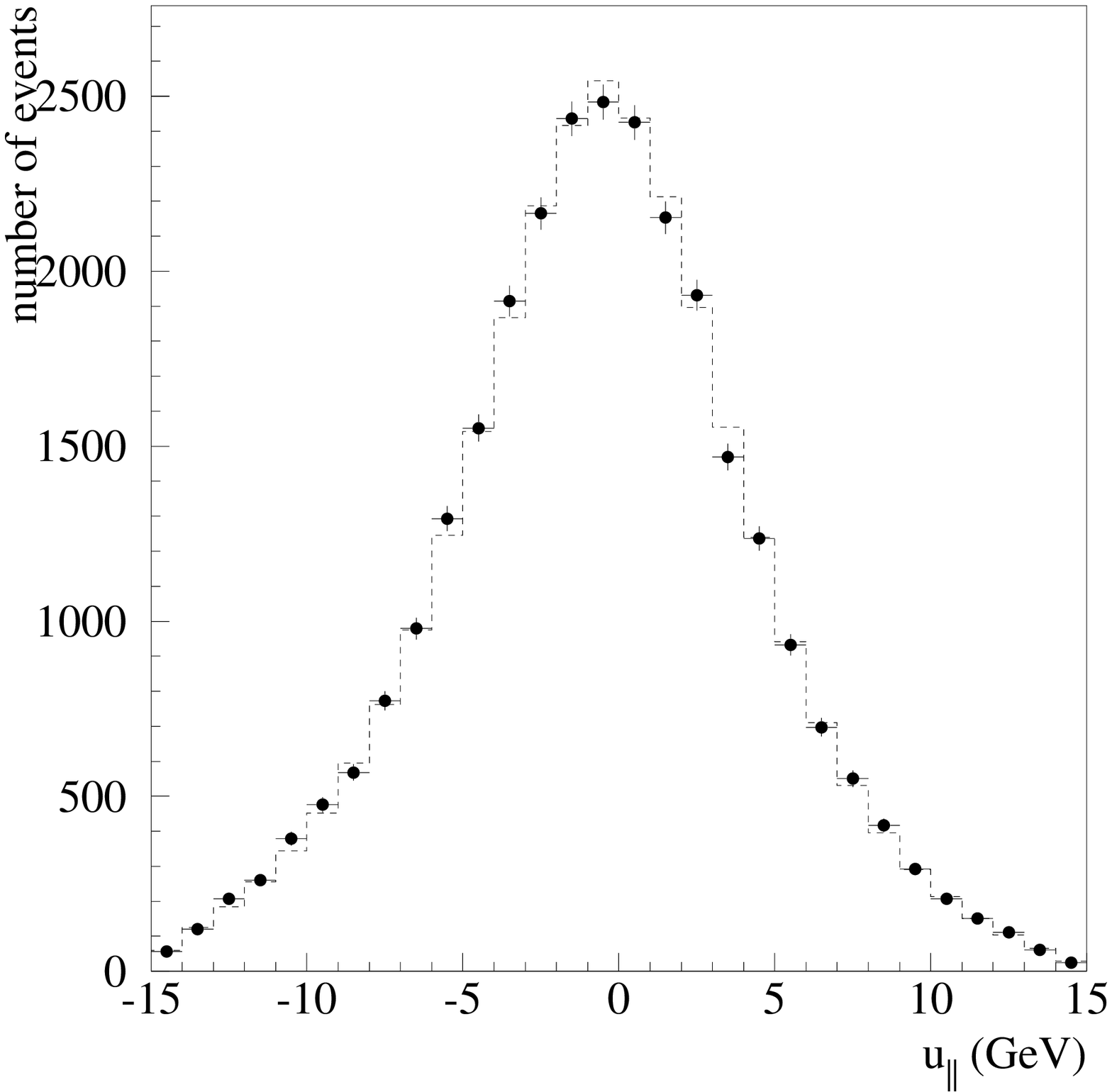}&
    \epsfxsize = 7.0cm \epsffile{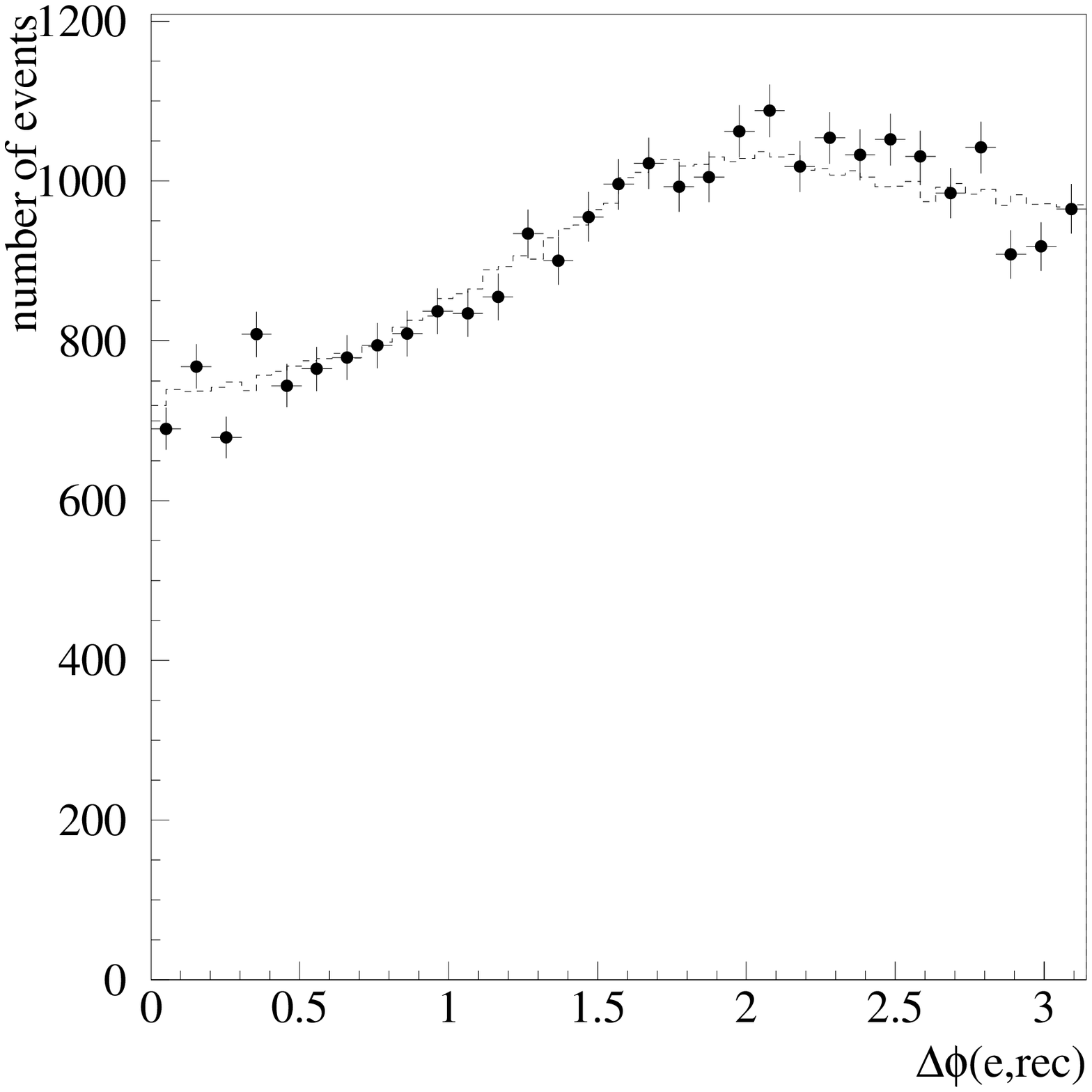}\\
  \end{tabular}
 \end{center}
  \caption{
    (a) Comparison of the $u_{\parallel}$ distribution from 
        $W\rightarrow e\nu$ events (points)
        and the MC simulation (histogram);
    (b) Comparison of the angle between the recoil and the electron 
         in the transverse plane 
        from $W\rightarrow e\nu$ events (points) and the 
        MC simulation (histogram).}
  \label{FIG:w2}
\end{figure}

\begin{figure}[h]
 \begin{center}
  \begin{tabular}{cc}
  \multicolumn{2}{c}{ \epsfxsize = 10.0cm \epsffile{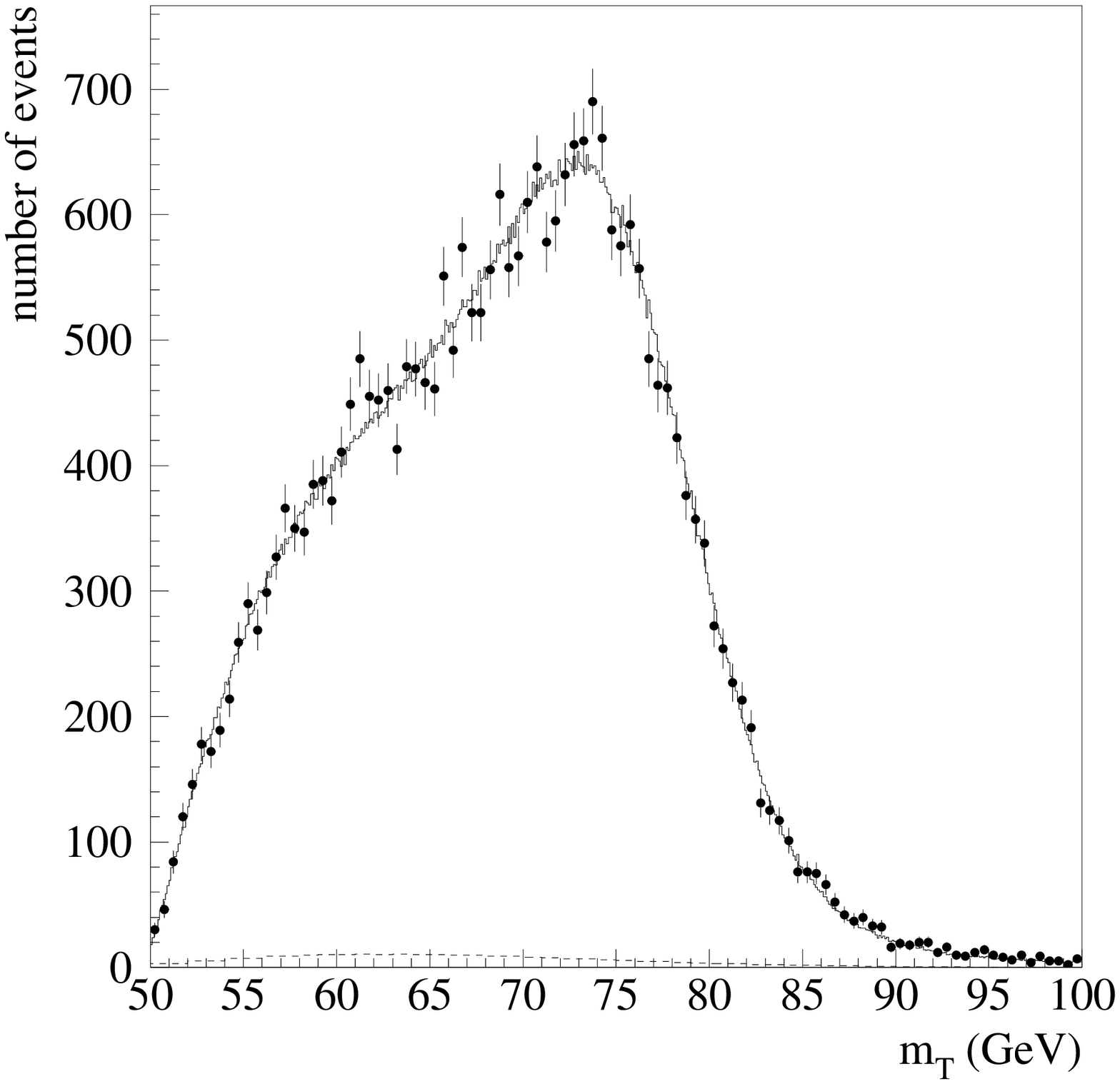}} \\
    \epsfxsize = 7.5cm \epsffile{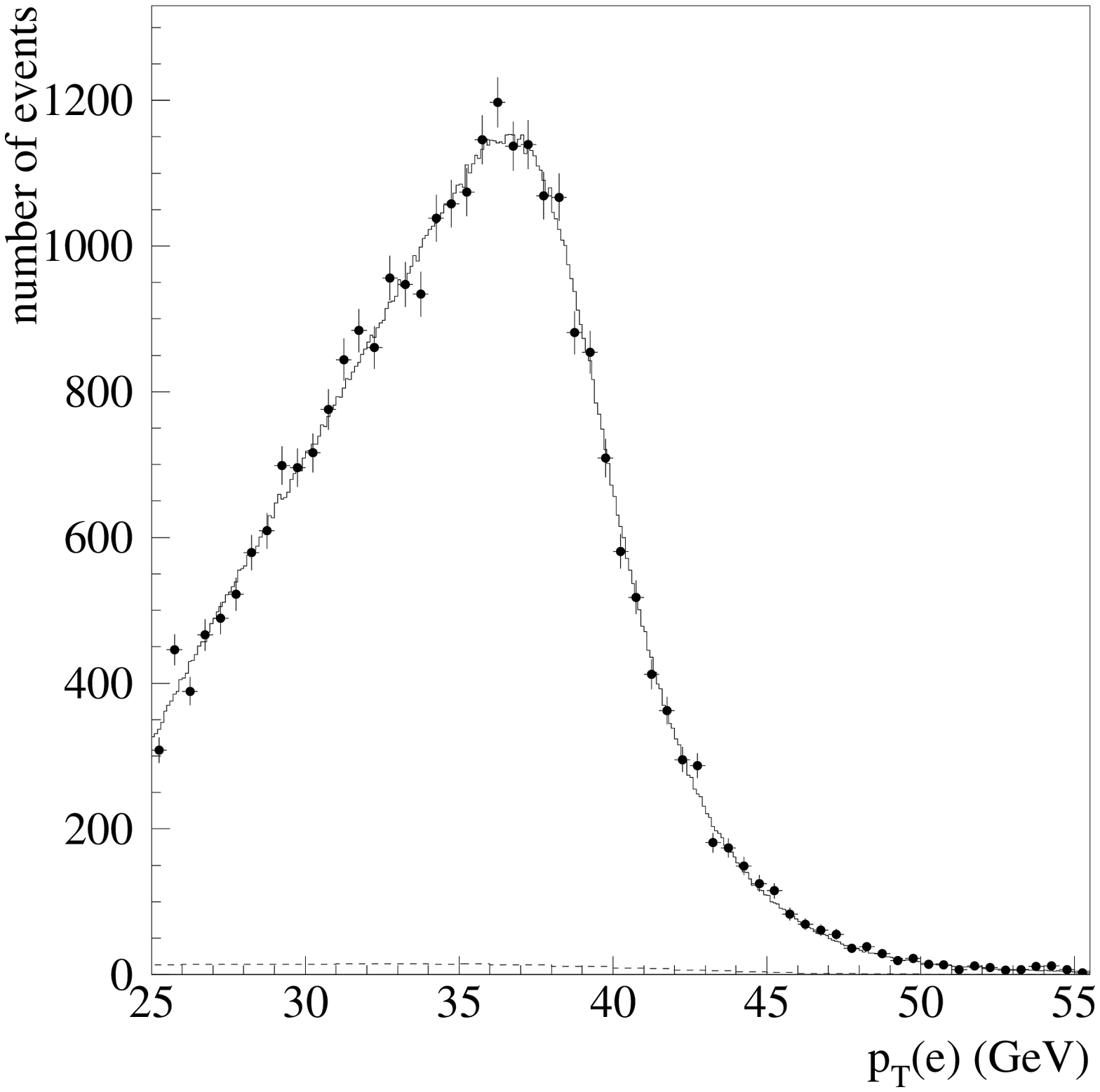}&
    \epsfxsize = 7.5cm \epsffile{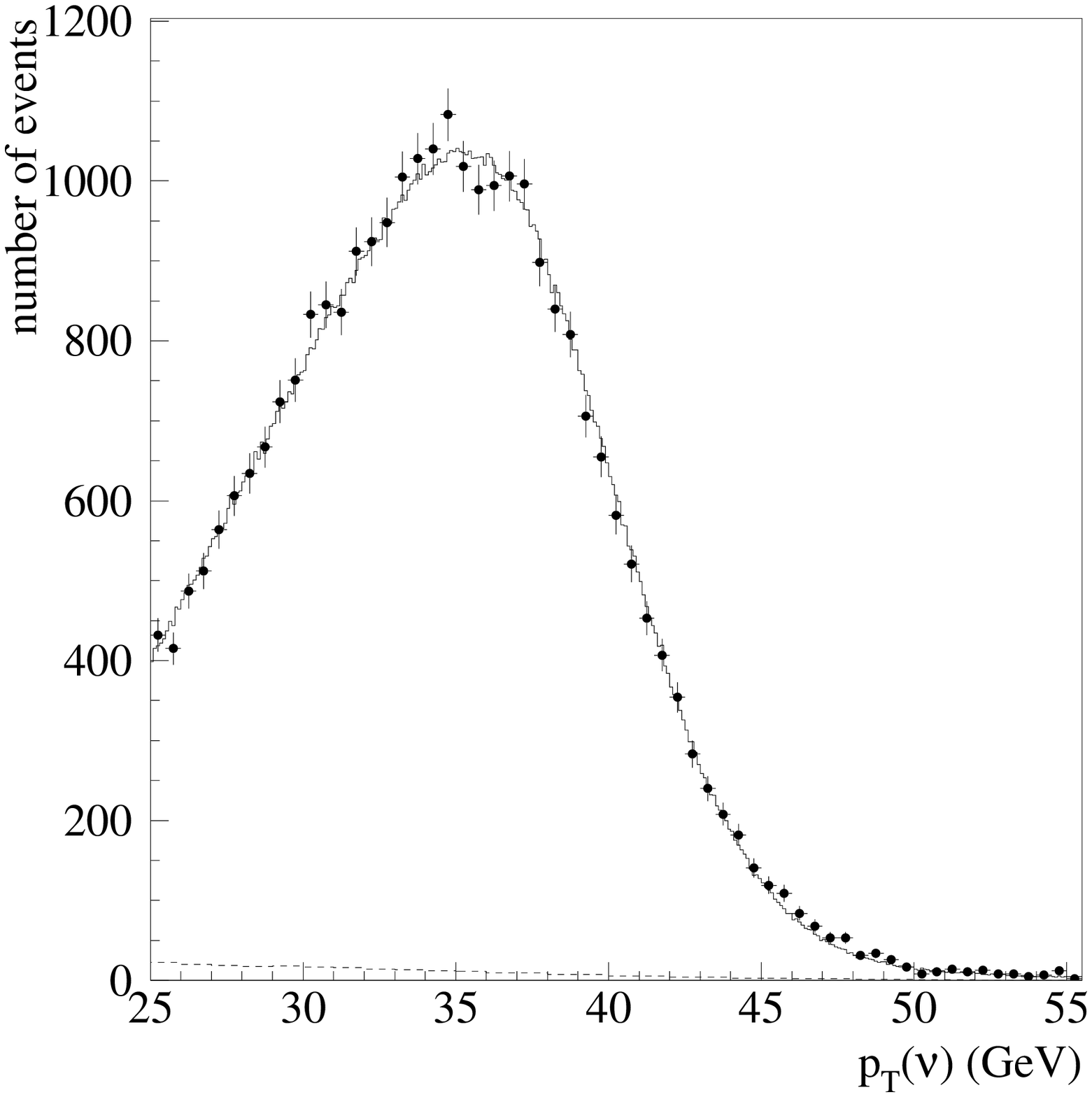}\\
  \end{tabular}
 \end{center}
 \caption{
   (a) The transverse mass distribution,
   (b) the electron transverse momentum distribution,
   and 
   (c) the neutrino transverse momentum distribution are
    shown for $W$ events (points).
    The best fits of the MC simulation (histograms) are also shown.}
 \label{FIG:w3}
\end{figure}               

Previous results from the run 1A data sample
  have been published\cite{wmass1a} and yielded a value of 
  $M_W=80.350 \pm 0.270~{\rm GeV/c^2}$.         
In the analysis presented here, the preliminary 
  measurement of $M_W$ from 
  the run 1B data sample is presented, using a calorimeter-based measurement.
The calorimeter is not calibrated independently to the precision needed
  and therefore the ratio of the $W$  to $Z$ masses was measured and
  then scaled to the  precisely known ($< 0.01\%$) $Z$ mass\cite{zmass}.
Many systematic errors cancel in this ratio.
                        
Experimentally,  the remnants of the interaction
  $p\bar p \rightarrow W(\rightarrow e\nu) + X$ are detected.
Here $X$ is due to the recoil $(rec)$ to the $W$ plus the  underlying event.
The energy of the electron  and the  $\vmet$ were measured.
The $\vmet= -\vec p_T(rec) - \vec p_T(e) =  \vec p_T(W) - \vec p_T(e)$ and is
  identified with the neutrino transverse momentum $\vec p_T (\nu)$ but differs
  from $\vec p_T (\nu)$ because of the presence of the underlying event.

\begin{figure}[h]
 \begin{center}
  \begin{tabular}{c}
    \epsfxsize = 10cm  \epsffile{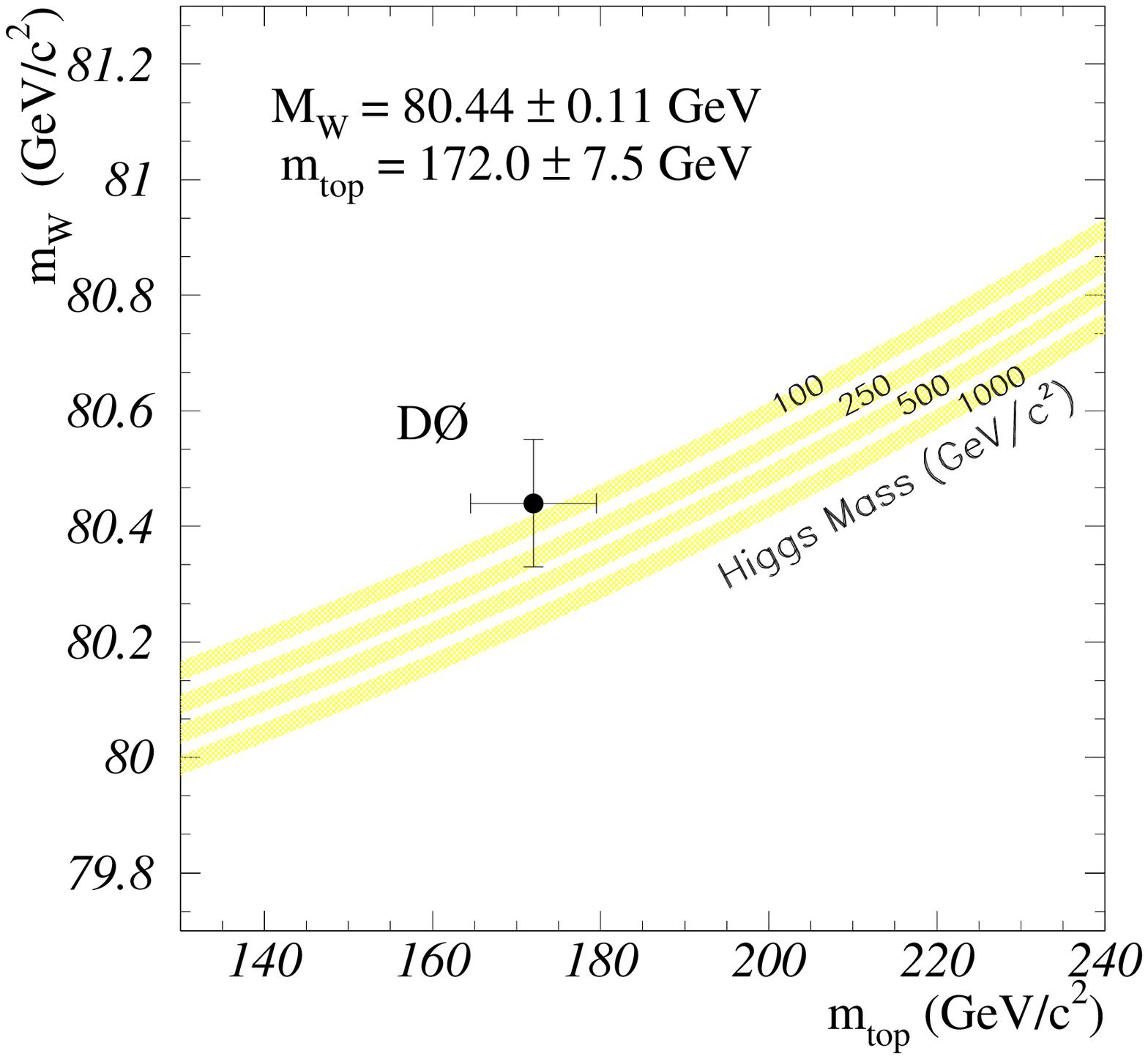}
  \end{tabular}
 \end{center}
  \caption{
  The D\O\ determination of the mass of the top quark is shown versus the
  measured $W$ mass.
  The SM prediction (see text) for various 
  assumptions of the Higgs boson mass is
  indicated by the bands.}
  \label{FIG:w4}
\end{figure}

Because the longitudinal momentum of the $\nu$ is not measured,
  the $W$  invariant mass cannot be constructed.
Instead  the distribution in transverse mass
  $M_T(W) = \sqrt{2p_T(e) p_T(\nu) - 2\vec p_T(e)\cdot \vec p_T(\nu) }$
  is used to obtain the $W$ mass.
For $Z$ decays, the energies of both electrons are measured and
  the invariant mass is reconstructed.

The $W\rightarrow e\nu$ events were selected by requiring an isolated
  electron with $E_T>25 ~\rm GeV$,  $p_T(W) < 15~\rm GeV/c$ 
  and  $\met >25~\rm GeV$.
The $Z\rightarrow ee$ events were selected  by requiring two isolated
   electrons each with $E_T>25 ~\rm GeV$,
   and $70 < M_Z < 110 ~{\rm  GeV/c^2}$.
Electrons were required to be in the region $|\eta|<1.0$.
There were 28323 $W$ events and 2179 $Z$ events in this sample.
The electron polar angle was determined from the
  shower centroid of the energy  cluster in the electromagnetic (EM)
  calorimeter and  the center-of-gravity of the corresponding track.
The uncertainty in determining this angle results in an uncertainty
  of $\pm 28 ~\rm MeV/c^2$ in  $M_W$.

The mass of the $W$ is determined by a maximum likelihood fit of the
  measured $M_T(W)$ distribution to Monte Carlo (MC) distributions which
  were generated for 21 different values of $M_W$ in $100 ~{\rm MeV}$ steps.
This fast MC simulation uses a theoretical calculation for the 
   $W$ production and
  decay and a parameterized model for the detector response.
Kinematic cuts are placed on the MC quantities as done in the data.
All the parameters in the MC are set by $Z$ data and other  data samples.
Below is a discussion of the determination of the parameters in the $W$ Monte
  Carlo.
The $Z$ data are treated in an analagous fashion.
Systematic errors are set using large statistics samples of MC data and
  varying the parameter within its errors and are discussed throughout.
               
The $W$ production is modelled by the double differential
  cross section in $p_T(W)$ and rapidity, $y$, 
  calculated at  next-to-leading order by Ladinsky and Yuan\cite{ladinskyyuan}
  and using the   MRSA\cite{mrsa} pdf.
The $W$ resonance is generated by a relativistic Breit-Wigner,
  incorporating the  mass dependence of the parton momentum distribution:
    \begin{eqnarray}
     \frac{d\sigma}{dm} \sim \frac{e^{-\beta\cdot m}}{m}\cdot \frac{m^2}
     {(m^2 - M_W^2)^2 + \frac{m^4\Gamma^2}{M_W^2}}. \nonumber
    \end{eqnarray}
The angular decay products are generated at ${\cal O}(\alpha_s)$, 
  allowing $p_T(W) > 0$,  in the $W$ rest frame.
This angular decay is of the form\cite{mirkes}
           $$ {{d\sigma}\over {d\cos \theta}} \sim
           ( 1 + \alpha_1 \cos \theta + \alpha_2 \cos^2 \theta) $$
  where $\alpha_{1,2}=\alpha_{1,2}(p_T(W))$\cite{mirkes}.
Radiative decays ($W\rightarrow e\nu\gamma$) are  generated according
  to Berends and Kleiss\cite{berendskleiss}.
Events in which $W\rightarrow \tau\nu \rightarrow e\nu{\bar \nu}\nu$
  are indistinguishable from $W\rightarrow e\nu$ decays and are
  therefore modelled in the simulation, including the polarization of
  the $\tau$ in the decay angular distribution.
The decay products are then boosted to the laboratory frame.
At this point, the values of  $p_T(e)$ and $p_T(W)$ have been generated 
  and $p_T(\nu)$ is calculated.
The effects of the detector and underlying event are now  modelled.
                                                                
The EM (electron) calorimeter energy scale of the calorimeter
  was determined using
  $J/\psi\rightarrow ee$, 
   $\pi^0\rightarrow \gamma\gamma\rightarrow e^+e^-e^+e^-$, and
  $Z\rightarrow ee$ events.
From test beam studies, it was determined that
  a linear relationship between the
  true and measured energies could be assumed:
  $ E_{\rm meas} = \alpha E_{\rm true} + \delta$.
This gives a relation
 $M_{meas} = \alpha M_{true} + \delta f$ between the measured and 
 true mass values, keeping terms to first order in $\delta$ only.
The variable $f = {2(E_1 + E_2)\over M} \sin^2 {\gamma\over 2} $ 
  depends on the event decay topology.
Since the ratio of $M_W$ to $M_Z$ is actually measured, one finds
  $$ \left({M_W \over {M_Z}}\right)^{meas} =
               \left({M_W \over {M_Z}}\right)^{true}
               \left[1 + {f\delta\over \alpha} \cdot
                   { (M_Z-M_W) \over{M_W M_Z }}   \right].$$
We note that to first order the measured ratio is insensitive to the 
  EM energy scale, if $\delta$ is small, and that the error on the
  measured ratio due to the uncertainty in $\delta$ is suppressed.

Figure \ref{FIG:w1} shows the mass spectra for
  the $\pi^0,J/\psi$ and $Z$ data samples.
The allowed ranges for $\alpha$ and $\delta$ are shown
  in Fig. \ref{FIG:w1}d for each data sample.
The overlap region is the $1\sigma$ contour from all three
  data samples.
The scale $\alpha$ is fixed by the $Z$ data.
The value of $\delta$ is constrained by the
  $J/ \psi$ and $\pi^0$ data,  essentially independent of $\alpha$.
Allowing a quadratic term in the energy response,
  to account for nonlinear responses  at low energies as measured
  at the test beam,  leads to the systematic error on $\delta$.
The allowed values determined for $\alpha$ and $\delta$ are
   $ \alpha = 0.95329 \pm 0.00077$ and
   $  \delta = -0.160 \pm 0.016(stat.) ^{~+.060}_{~ -.210}(syst.) ~\rm GeV$.
The error in the EM energy scale introduces an uncertainty in $M_W$
  of $\pm 65 {\rm ~MeV/c^2}$ and is dominated by the statistical error
  in determining the $Z$ mass.

The EM energy resolution is parameterized
 as ${\sigma / E} =
       \sqrt{C^2 + ({S/ {\sqrt {E_T}}})^2 + ({N/ E})^2 }$
 for the central calorimeter.
Test beam data are used to set the sampling term,
  $S = 0.13 ~\rm {(GeV^{1/2})}$, and  the noise term, $N=0.4~\rm {GeV}$.
By constraining the width of the $Z$ invariant mass distribution in the
  MC to that from the data, the constant term is set to
  $C = (1.15~ {^{+0.27}_{-0.36}})$.
The uncertainty in the energy resolution leads to an uncertainty of
  $\pm 23~\rm MeV/c^2$ in $M_W$.

The hadronic (recoil) energy scale of the calorimeter
  is determined relative to the EM energy scale by
  using $Z$ events and measuring the transverse
  momenta of the $Z$ from both the recoil or the two electrons.
The $p_T{\rm -balance}$ is constructed:
       $$p_T{\rm -balance} \equiv
             [\vec p_T(ee) + \vec p_T(rec)]\cdot {\hat \eta}$$
  where $\hat \eta$ is defined as the bisector of the 
  two electrons.
From studies using {\footnotesize{HERWIG}}\cite{herwig} and 
  {\footnotesize{GEANT}}\cite{geant},   it was determined that
  the recoil response could be written as a function of
  the EM response: $p_T(rec) = {\cal R}_{rec} p_T(ee)$ 
  with ${\cal R}_{rec} = \alpha_{rec} + \beta_{rec} {\rm log} ~p_T(W)$.
To ensure an equivalent event topology, $Z$ events in which one electron
  is in the forward region are included in this study.
Comparing data to MC in a plot of $p_T{\rm -balance}$ versus
  $\vec p_T(ee)\cdot {\hat \eta}$,
  the recoil response parameters are determined to be
  $\alpha_{rec} = 0.69 \pm 0.06$ and $\beta_{rec} = 0.04 \pm 0.02$.
The  uncertainty in the recoil scale leads to an uncertainty
  of $\pm 23 ~\rm {MeV/c^2}$ in $M_W$.

The recoil (hadronic) energy resolution is determined by
  modelling both components of the recoil to the $W$:
  $ \vec p_T(rec)^{meas} =  
  {{\cal R}_{rec}}\vec p_T(rec) + 
  \alpha_{mb} \cdot \vec U_{mb}(tot) -  U(\hat e)$.
The first component is the ``hard" component due to the initial $p_T$ of the
  boson.
It is smeared using a Gaussian of width  
  $\sigma_{rec} = s_{rec} \sqrt{p_T(rec)}$.
The second component is the ``soft" component due to the underlying
  event and  is modelled by a  minimum-bias event obtained from the data.
In selecting the minimum-bias events to use, the luminosity distribution of
  the $W$ event sample is modelled.
The quantity $\vec U_{mb} $ is the total $\vec E_T$ of minimum-bias event
  and $\alpha_{mb}$ is a scale factor.  
The amount of underlying event in the electron direction,  $U(\hat e)$,
  is subtracted from the recoil and added onto the electron momentum.
Using the width of the $p_T{\rm -balance}$ distribution (to which the
  energy calibration has been applied), the values of $s_{rec}$ and 
  $\alpha_{mb}$   are constrained.
The measured values are $s_{rec}= 0.49 \pm 0.14$ and 
  $\alpha_{mb}= 1.03\pm 0.03$ and their errors and lead to an uncertainty of 
  of $\pm 33 ~\rm {MeV/c^2}$ in $M_W$.

Selection biases due to radiative decays and the
  amount of  recoil energy in the electron direction 
  and trigger efficiences are modelled in the MC simulation.
The uncertainty in $M_W$ due to the modelling of these
  efficiencies and biases is negligible in the fit to
  the $M_T(W)$ spectrum.

Backgrounds to the $W$ event sample are included in
  the fitting procedure by including the shape and fraction of background
  events.
The largest source of background in the $W$ sample is due to QCD multijet
  production in which there is a jet is mis-identified as an electron and
  $\met$ due to energy fluctuations.
This background contributes $1.4\pm 0.2 \%$ to the $W$ sample.
The other background considered is $Z\rightarrow ee$ events 
  where one electron  is not identified.
This background contributes $0.33\pm 0.06 \%$ to the $W$ sample.
The uncertainty in size and shape of the backgrounds gives an uncertainty in
  $M_W$ of $\pm 12~\rm {MeV/c^2}$.
All other sources of background are negligible.

The last systematic error to consider is that  due to the modelling 
  of the $W$  production.
This uncertainty is due to the correlated uncertainties in the $p_T(W)$ 
  spectrum and the pdf's.
There are three phenomenological parameters in the production model 
  calculation 
  ($g_1,g_2,g_3$)\cite{ladinskyyuan} and  the largest sensitivity of 
  the $p_T$ spectrum  is to the  $g_2$ parameter.
To constrain the production model, the $g_1$ and $g_3$ parameters are fixed to
  their nominal values and the value of $g_2$ is constrained by
  the $p_T(Z)$ distribution from the data.
Then the dependence of $M_W$ on the pdf used in the theoretical calculation
  is measured from the difference in $M_W$ from the nominal pdf (MRSA) as
  seen in Table \ref{TABLEw1}.
For each pdf, the theoretical calculation uses the value 
  of $g_2$ constrained  by the data for that case.
The uncertainties on the measured $M_W$ due to the value of $g_2$ and
  the pdf used are  $\pm 5~\rm {MeV/c^2}$ and $\pm 21~\rm {MeV/c^2}$, 
  respectively.
Errors on $M_W$ are also ascribed to uncertainties in 
  the value of $\Gamma_W \rightarrow \pm$ 9 MeV,
  the parton luminosity parameter $\rightarrow\pm$ 10 MeV, and
  the modelling of radiative decays $\rightarrow\pm$ 20 MeV.
The total uncertainty on $M_W$ due to the production 
  model is $\sigma (M_W) = \pm 34 {\rm ~MeV}$.

 \begin{table}[h]
 \begin{center}
   \begin{tabular}{rcr}
      pdf & constrained $g_2$ & $\Delta M_W~{\rm MeV}$ \\ \hline
      MRSA   & $0.59$ & -   \\
      MRSD-  & $0.61$ & +20 \\
      CTEQ3M & $0.54$ & +5  \\
      CTEQ2M & $0.61$ & -21 \\ 
\end{tabular}
 \caption{The variation of the measured $W$ mass when using 
        different pdf's in the production model.  Each
         theoretical calculation uses the constrained 
         value of $g_2$ for that pdf.}
 \label{TABLEw1}
\end{center}
\end{table}

A measure of how accurately the MC describes the data is shown
   in Fig. \ref{FIG:w2}.
The quantity  $u_{\parallel} = \vec p_T(rec)\cdot \hat e$,
  which is the hadronic energy in electron direction, is
  shown in  Fig. \ref{FIG:w2}a.
A bias in $u_{\parallel}$ directly affects the $M_T(W)$ spectrum
  and it is also very  sensitive to the recoil resolution.
Another sensitive quantity is the difference in the azimuthal angle, 
   $\Delta\phi$,
   between the electron and the recoil and is shown in Fig. \ref{FIG:w2}b.
Excellent agreement between the data and MC  simulation is obtained.

\begin{table}[h]
\begin{center}
\begin{tabular}{lc}
Source & \multicolumn{1}{c}{$\sigma_{M(W)}$ in  $\rm {MeV/c^2}$ }\\
Statistical (W events)   & 69  \\ \hline
Statistical (Z events)   & 65  \\ \hline
% ***  detector modeling
 \quad  Non-Uniform energy response ($\eta$) & 10 \\
 \quad  Electron Angle Calibration           & 28 \\
 \quad  Electron Energy Resolution           & 23 \\
 \quad  Electron Energy Linearity            & 20 \\
 \quad  Electron Underlying Event           & 16 \\
%{{   Selection Biases/Efficiencies }}       & - \\
 \quad  Hadronic Energy Scale               & 20 \\
 \quad  Hadronic Resolution                  & 33 \\
% *** theory
 \quad  $P_T(W)$ Spectrum                    & 5 \\
 \quad pdf                                 & 21 \\
 \quad  parton luminosity                    & 10 \\
 \quad  $W$ Width                           & 9 \\
 \quad  Radiative Decays                     & 20 \\
% *** backgrounds
 \quad  QCD background                       & 11 \\
 \quad  Z background                         &  5 \\
Systematic Total                            & 70  \\ \hline \hline
Total                                       & 118 \\ 
\end{tabular}
\caption{Summary of errors on the $W$ mass measurement.}
\label{TABLEw2}
\end{center}
\end{table}

The $M_T(W)$ distribution from the data is shown in Fig. \ref{FIG:w3}a 
   together with the
  distribution from the best fit value of $M_W$ from the 
   Monte Carlo simulation.
The data are fit over a region $60$ to $ 90 ~\rm {GeV/c^2}$ and the preliminary
  value of the $W$ mass determined is
  $M_W = 80.450 \pm 0.070(stat.)
                \pm 0.065(scale)
                \pm 0.070(syst.) ~ {\rm GeV/c^2}$, giving a
  total error of $\pm 118 ~{\rm MeV/c^2}$.
The errors on the $W$ mass are detailed in Table \ref{TABLEw2}.

As consistency checks, the $p_T(e)$ and $p_T(\nu)$ spectra are also
  fit to determine $M_W$ as shown in Figs. \ref{FIG:w3}b and \ref{FIG:w3}c.
The fit to the $p_T(e)$ spectrum gives
  $ M_W = 80.49 \pm 0.14 ~{\rm GeV/c^2}$
  and the $p_T(\nu)$ fit gives
  $M_W = 80.42 \pm 0.18(stat.)  ~  {\rm GeV/c^2}$
   with the fitting region from  $30$ to $ 50 ~\rm {GeV/c}$ in both cases.

In summary, the measured $W$ masses from the D\O\ data sample
  of  $W\rightarrow e\nu$ decays with the $e$ in the central
  $\eta$ region are
  $ M_W = 80.35 \pm 0.27 ~{\rm GeV/c^2}$ (run 1A) and
  $ M_W = 80.45 \pm 0.12 ~{\rm GeV/c^2}$ (run 1B, preliminary).
Combining these results and taking into account the correlated errors 
  gives a D\O\ combined value of  $M_W = 80.44 \pm 0.11 ~\rm {GeV/c^2}$.

Combining the new D\O\ result with other measurements\cite{mwua2,mwcdf}
  from hadron collider experiments gives a new preliminary
  hadron collider average of $M_W = 80.41 \pm 0.09 ~\rm {GeV/c^2}$.

The constraints placed on the Higgs mass can be seen in Fig. \ref{FIG:w4},
  which shows the measured values of $M_W$ versus $m_{top}$\cite{mtopd0} 
  compared to the Standard Model prediction\cite{smhiggs} for different values
  of the $M_H$ in the $m_{top}-M_W$ plane.

%==========================================================================

\section{Conclusions}

In conclusion, D\O\ has collected   $\approx 100 {\rm pb}^{-1}$ of data
     from run 1 of the Tevatron and preliminary results on $W$ boson
     properties are found to be in agreement with the Standard Model.
The $W$ and $Z$ cross sections are measured in the
  $e,\mu,\tau$ decay modes.
The $W$ width is measured to be $\Gamma_W = 2.159 \pm 0.092 ~{\rm GeV}$.
We confirm $e-\tau$ universality in $W$ decays with the measurement
  $g_{\tau}^W/ g_e^W = 1.004 \pm 0.032$.
The run 1 combined D\O\ $W$ mass,
   $M_W = 80.44 \pm 0.11~{\rm GeV/c^2} $, is currently the most accurate
  direct measurement.


\begin{references}
\bibitem{smpred}
  We use ref. \cite{wzsmpred} 
  with the CTEQ2M pdf, $M_W = 80.23 \pm 0.18$ GeV, $M_Z = 91.19$ GeV
  and $\sin^2\theta_W = 0.2259$.
\end{references}
\end{document}